\documentclass[aps,prb,reprint,floatfix,superscriptaddress]{revtex4-1}
\usepackage[english]{babel}
\usepackage[utf8]{inputenc}
\usepackage[T1]{fontenc}
\usepackage{bbold}
\usepackage{epsfig}
\usepackage{amsmath}
\usepackage{hyphenat}
\usepackage{hyperref}
\usepackage{placeins}
\usepackage[justification=raggedright]{caption}
\usepackage{subcaption}	
\usepackage[colorinlistoftodos]{todonotes}

\begin{document}

\title{Stability of the X-Y-phase of the two-dimensional $C_4$ point group insulator}
\author{Bart de Leeuw}
\affiliation{Institute for Theoretical Physics and Center for Extreme Matter and Emergent Phenomena,
Utrecht University, Leuvenlaan 4, 3584 CE Utrecht, The Netherlands}
\affiliation{Mathematical Institute, Utrecht University, Budapestlaan 6, 3584 CD Utrecht}
\author{Carolin K\"uppersbusch}
\affiliation{Institute for Theoretical Physics and Center for Extreme Matter and Emergent Phenomena,
Utrecht University, Leuvenlaan 4, 3584 CE Utrecht, The Netherlands}
\affiliation{Institut für Theoretische Physik, Universität zu Köln, Zülpicher Straße 77, 50937 Köln, Germany}
\author{Vladimir Juri\v ci\' c}
\affiliation{Institute for Theoretical Physics and Center for Extreme Matter and Emergent Phenomena,
Utrecht University, Leuvenlaan 4, 3584 CE Utrecht, The Netherlands}
\author{Lars Fritz}
\affiliation{Institute for Theoretical Physics and Center for Extreme Matter and Emergent Phenomena,
Utrecht University, Leuvenlaan 4, 3584 CE Utrecht, The Netherlands}

\begin{abstract}
Noninteracting insulating electronic states of matter can be classified according to their symmetries in terms of topological invariants which can be related to effective surface theories. These effective surface theories are in turn topologically protected against the effects of disorder. Topological crystalline insulators are, on the other hand, trivial in the sense of the above classification but still possess surface modes. In this work we consider an extension of the Bernevig-Hughes-Zhang model that describes a point group insulator. We explicitly show that the surface properties of this state can be as robust as in topologically nontrivial insulators, but only if the $S_z$-component of the spin is conserved. However, in the presence of Rashba spin-orbit coupling this protection vanishes and the surface states localize, even if the crystalline symmetries are intact on average.
\end{abstract}
\maketitle

\section{Introduction}

Soon after the prediction~\cite{kane2005,kane2005a,bernevigzhang2006} and subsequent discovery of the quantum spin Hall insulator~\cite{koning2006} as a novel state of electronic matter with properties protected by topology, it became clear that this state fits into an even grander scheme.~\cite{moore2007,fukane2006,fukane2007a,fukane2007b} By now many more topological phases have been identified both theoretically and experimentally.~\cite{hsieh2008,hsieh2009,xia2009,zhang2009,chen2009,hasan-kane-review,qi-zhang-review} The classification of noninteracting electronic insulating states with topological order was one of the milestones of theoretical condensed matter physics in the last decade.~\cite{schnyder2008,kitaev2009,Qi2008} Within the so-called 'tenfold periodic table' states of matter with topological order are classified according to the presence or absence of symmetries of the underlying systems,~\cite{schnyderNJP2010} such as time-reversal, particle-hole, and chiral.

One of the prominent features of topological electronic systems is the existence of exotic gapless edge or surface states. In particular, these boundary modes can realize one-dimensional chiral or single two-dimensional Dirac fermions, usually ruled out by the fermion doubling theorem (at least if all the symmetries are preserved).
Importantly, in the presence of disorder these states can be protected against localization which leads to unusually robust transport properties. An instructive point of view on the existence of these robust boundary states is that the tenfold periodic table does not require spatial symmetries such as translations or rotations to be present, which is particularly apparent in the classification scheme using non-linear $\sigma$ models.~\cite{schnyder2008} This implies that although the calculation of topological invariants can in general most easily be accomplished in clean band-insulators, in principle there is no need to have a well-defined crystal momentum for doing so.
On the other hand, electronic states trivial according to the tenfold classification but still featuring edge or surface modes  have been recently identified once crystalline symmetries, such as translations, rotations, reflections and inversions, were taken into account.\cite{fu2011,slager2013,chiu2013,morimoto2013,shiozaki2014} These states are conventionally referred to as topological crystalline insulators~\cite{fu2011} (TCI). Interestingly, these states, predicted to be realized in Sn- and Pb-based compounds, \cite{fuNatComm2012} have been reported to be observed first in Refs.~\onlinecite{ando2012,xu2012,story2012}, and subsequently have been  experimentally studied in Refs.\ \onlinecite{tanaka2013,okada2014,zeljkovic2014}. 
An urging question is therefore whether the boundary states in topological phases where crystal symmetries allow to define topological invariants can enjoy similar protection against the effects of disorder as they do in tenfold-wise topologically nontrivial insulators.

To answer this question at least for one concrete example, we investigate the transport properties of an extension of the well-known Bernevig-Hughes-Zhang (BHZ) model that was recently introduced.~\cite{slager2013}, to which we refer as a point group insulator (PGI) in the following. There, it was identified that a rotational symmetry leads to the existence of a state characterized by a trivial topological invariant in the tenfold sense, but non-trivial in a sense defined in Sec.\ II. Within this note we make a comparison of this PGI with a standard quantum spin Hall insulator (QSHI), for which the BHZ model was originally formulated.\cite{bernevigzhang2006} For the QSHI we find, as expected and well known, very robust transport properties with a quantized lead-to-lead conductance, both in presence and absence of disorder (as long as the disorder strength is smaller than the bulk-gap scale). Importantly, this property is also robust against breaking of the spin-rotational symmetry, introduced for instance by Rashba spin-orbit coupling, and is a direct consequence of time-reversal symmetry. For the PGI, on the other hand, we find that the conductance is only quantized and robust against disorder if the $S_z$-component of the spin is conserved. However, in the presence of Rashba spin-orbit coupling disorder localizes, in the Anderson sense, the boundary modes leading to a vanishing conductance, in agreement with the specific PGI state being trivial in the sense of the periodic table.
 We note here that our results concerning the stability of the edge modes in a PGI without Rashba spin-orbit coupling are in agreement with the findings recently reported in Refs.\ \onlinecite{jiang2014} and \onlinecite{ezawa2014}, but also show that the results obtained therein are fine tuned and non generic.

The paper is organized as follows. In Sec.\ II, we introduce the model and its phase digram, and in Sec.\ III, we present the results concerning the transport properties in both QSHI and PGI phases. Our conclusions are drawn in Sec.\ IV.

\section{Model, Topological Invariants and Phase Diagram}

The model we study was recently introduced in Ref.\ \onlinecite{slager2013} and represents an extension of the BHZ Hamiltonian that includes next-nearest neighbor hoppings. Its Bloch Hamiltonian has the generic form
\begin{eqnarray}\label{eq:H}
\mathcal{H}=\sum_{\bf{k}}\Psi^{\dagger}_{\bf{k}} \left( \begin{array}{cc} H ({\bf{k}}) & H_{{\rm{SO}}} ({\bf{k}}) \\  H^\dagger_{{\rm{SO}}} ({\bf{k}}) & H^*(-{\bf{k}}) \end{array} \right) \Psi^{\phantom{\dagger}}_{\bf{k}},
\end{eqnarray}
where
\begin{widetext}
\begin{eqnarray}
H({\bf{k}})={\bf{\tau}} \cdot {\bf{d}}({\bf{k}}), \quad {\rm{with}} \quad {\bf{d}}({\bf{k}})=\left(\begin{array}{c} \sin k_x+ \cos k_x \sin k_y \\ -\sin k_y +\cos k_y \sin k_x \\ M-2B \left[ 2-\cos k_x -\cos k_y \right] -4\tilde{B} \left[1- \cos k_x \cos k_y \right] \end{array} \right),
\end{eqnarray}
\end{widetext}
 and ${\bf{\tau}}=(\tau_x,\tau_y,\tau_z)$ is the vector composed of the Pauli matrices. The wave function is a four component object, and, motivated by the low-energy band-structure of HgTe/CdTe quantum wells, contains $s-$ and $p-$like orbitals with the respective spins $ \Psi_{\bf{k}}= \left(s_{\bf{k}}^\uparrow,p_{\bf{k}}^\uparrow,s_{\bf{k}}^\downarrow,p_{\bf{k}}^\downarrow \right)$. The parameter $M$ is related to the offset of the chemical potential between the $s-$ and $p-$like orbitals, while $B$ (${\tilde B}$) is proportional to the (next-)nearest-neighbor hopping between the orbitals of the same type, while both nearest-neighbor and next-nearest-neighbor hoppings between different orbitals are set to one.

We choose a version of Rashba spin-orbit coupling which is not symmetric in the orbitals $s$ and $p$\cite{rothe2010} but respects all other symmetries discussed below. The corresponding term $H_{\rm{SO}}$ in the Hamiltonian (\ref{eq:H}) reads
\begin{eqnarray}
H_{\rm{SO}}=R_0 \left(\begin{array} {cc} -i \sin k_x - \sin k_y & 0 \\ 0 & 0  \end{array} \right) \;.
\end{eqnarray}
The Hamiltonian (\ref{eq:H}) obeys time reversal symmetry, which is implemented via $\mathcal{T}=i \tau_0 \otimes \sigma_y K$, where $\sigma_y$ acts in spin space and $K$ denotes complex conjugation ($\tau_0$ is the identity in orbital space). This symmetry will be intact hereafter.
Additionally, the clean system has the discrete rotational $C_4$ symmetry about the axis orthogonal to the crystal plane,  represented by $\mathcal{R}=\frac{1}{2}\left( \tau_0 \left(1+i \right)+\tau_z \left(1-i \right)  \right)\otimes{\cal R}_{\rm spin}$, with ${\cal R}_{\rm spin}=\exp(i\pi\sigma_z/4)$. The invariance of the Hamiltonian (\ref{eq:H}) under this symmetry operation is  shown in the Appendix. However, this symmetry will be broken by disorder and the boundary of the sample.
\begin{center}
\begin{figure}
\includegraphics[width=0.48\textwidth]{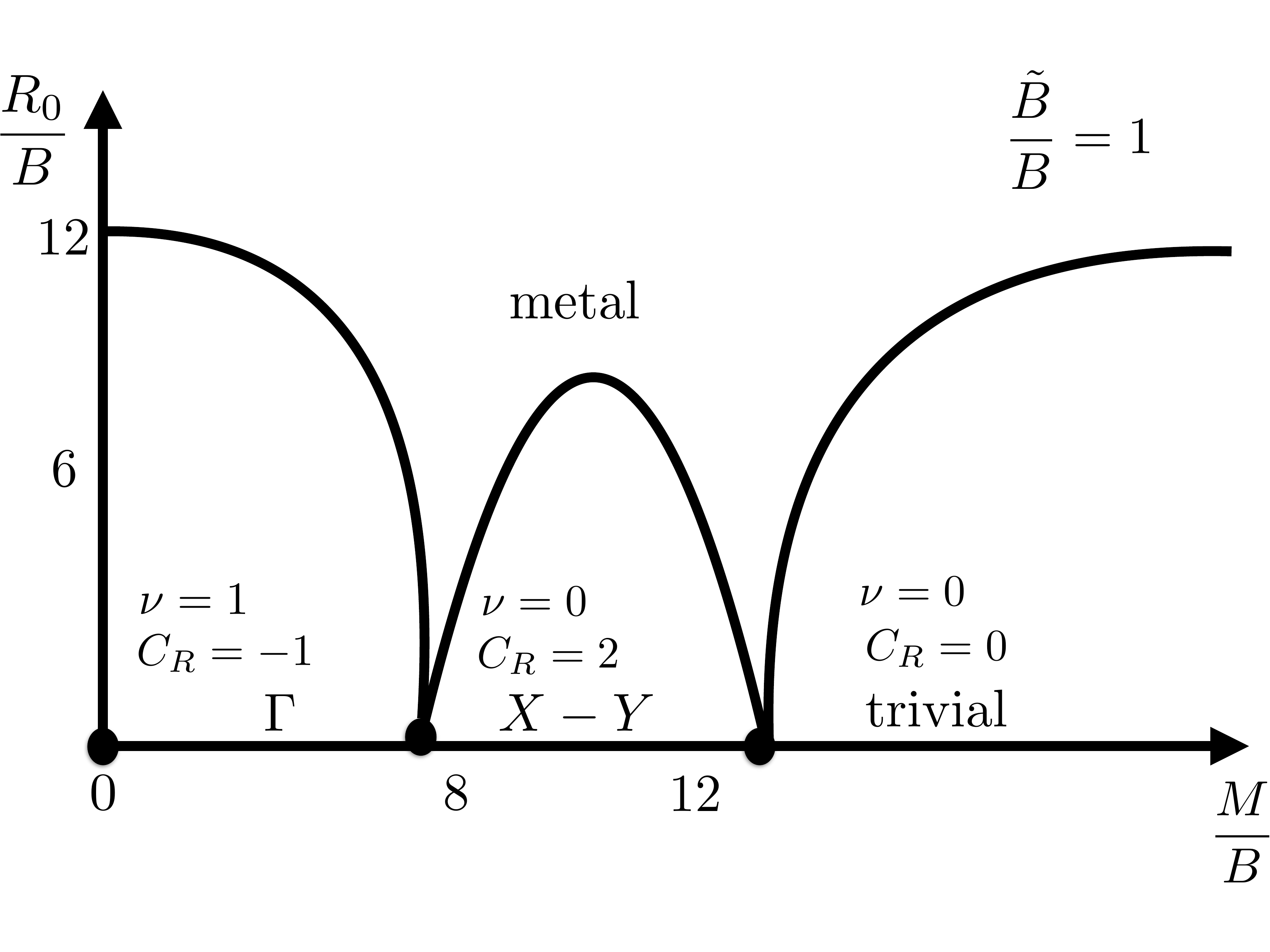}
\caption{Phase diagram of the clean system. The phase diagram without Rashba spin-orbit coupling is obtained from the calculation of the topological $C_4$ index $C_R$, and the corresponding $Z_2$ index $\nu$.} \label{fig:pd}
\end{figure}
\end{center}

\subsection{Topological Characterization of the phases on a square lattice}

Before we study the model \eqref{eq:H}, we discuss a topological characterization of the possible topological phases on the square lattice. 
In Ref.\ \onlinecite{slager2013} it was shown that in the presence of time-reversal symmetry, the distinct topological phases on a lattice with a given space group  symmetry can be characterized in terms of the band inversions at time-reversal invariant (TRI) momenta in the Brillouin zone (BZ). The band inversion is given in terms of the sign of the Pfaffian at TRI momenta. In the BHZ model the sign of the Pfaffian at a TRI momentum is equal to the sign of the mass term [$d_z ({\bf k})$ in Eq.\ \eqref{eq:H}] at that point. \cite{slager2013,ezawa2014} When these points are, in addition, related by the $C_4$ rotational symmetry, as is the case with $X$ and $Y$ points on the square lattice, the corresponding band inversions are connected to each other. If there is a band inversion at $X$ point, there is one at $Y$ point as well, and this is enforced precisely by the $C_4$ symmetry,\cite{slager2013} as it is the case in the  $X-Y$ phase, which is obtained in the model \eqref{eq:H}, as detailed below. Notice, however, that this phase has a trivial $Z_2$ topological invariant, which is also confirmed by our stability analysis. Finally, given the gauge of the Bloch states, the sign of the Pfaffian cannot be changed without closing the band gap, and thus $X-Y$ phase is topologically distinct from other possible phases on the square lattice, $\Gamma$ and $M$ phases, which both have nontrivial $Z_2$ invariant.\cite{slager2013}

An alternative way to show that the $X-Y$ phase is topologically distinct from the other topological phases on a square lattice was outlined by Ezawa in Ref.\ \onlinecite{ezawa2014}, where he derived the form of the mirror Chern number in terms of the signs of the Pfaffians at TRI points, Eq. (34) therein. Just by interchanging the mirror operator with the $C_4$ operator, which is allowed since the only ingredient necessary in this derivation is the commutation of the Bloch Hamiltonian with the point-group operation, \cite{footnote}  the topological invariant associated with $C_4$ symmetry is also given in terms of signs of the Pfaffians $\delta_{k_{i}}$ at TRI points $k_i$.  Therefore, the topological invariant  corresponding to $C_4$ symmetry is
\begin{equation}
C_R=\frac{1}{2} (\delta_\Gamma + \delta_M - 2\delta_X),
\end{equation}
since $\delta_X=\delta_Y$ due to $C_4$ symmetry. In the $X-Y$ phase, $\delta_X=\delta_Y=-1$, while $\delta_\Gamma=\delta_M=+1$, so $C_R=2$. On the other hand,  in the $\Gamma$ ($M$) phase, with $\delta_\Gamma=-1$ ($\delta_M=-1$),  $C_R=-1$, and therefore both $\Gamma$ and $M$ phases are topologically distinct from the $X-Y$ phase.

In the $X-Y$ phase, depending on the boundary, the two resulting edge modes lie at different TRI momenta, and this is important for the resulting physics of the edge states. For instance, when the cut is along one of the two principal crystallographic axes $x$ or $y$, the two edge states originate from the Dirac cones located  at the TRI momenta $k_{\rm edge}=0$ and $k_{\rm edge}=\pi$ in the boundary BZ. On the other hand, when the cut is along the diagonal, the two Dirac cones are both located at a non-TRI momentum $k_{\rm edge}=\pi/\sqrt{2}$ . The types of the edges are completely analogous to the surface states discussed in Ref.\ \onlinecite{fu-surface} in the context of three-dimensional TCIs on the rocksalt lattice with the bulk band inversion at the four $L$ points in the BZ. The surface perpendicular to the $(111)$ crystallographic direction in this 3D TCI corresponds to an edge along one of the principal axes in the $X-Y$ phase, since the corresponding edge/surface Dirac cones are separated by a finite momentum, and therefore a finite momentum transfer is needed to mix them. On the other hand,  the $(001)$ surface in the 3D TCI is analogous to the edge along one of the diagonals in the $X-Y$ phase, since in both cases the surface/edge Dirac cones lie at the same momentum, and can therefore hybridize.

These general arguments imply that  upon adiabatically switching on the Rashba-coupling in the $X-Y$ phase, provided that the bulk gap remains open, the two Dirac points cannot be mixed and consequently edge states are stable, as long as there is no perturbation providing the momentum transfer required to mix the two Dirac cones. As mentioned before, there is one special situation in which this is not true, however, and that is when the two Dirac cones corresponding to the two Kramers pairs of edge states coincide in the edge BZ, corresponding to a diagonal cut, as discussed above. In that situation a gap will open due to Rashba spin-orbit coupling.
The most natural candidate to generically mix the Dirac cones in the presence of Rashba spin-orbit coupling is scalar disorder, which we consider in the paper.

\subsection{Phase diagram}

The phase diagram for the model \eqref{eq:H} is shown for $\tilde{B}/B=1$ as a function of $M/B$ and $R_0/B$ in Fig.~\ref{fig:pd}. The phase diagram for $R_0=0$ and all values of $\tilde{B}/B$ was introduced recently.~\cite{slager2013}
If $R_0=0$, the Hamiltonian explicitly conserves the $S_z$-component of the spin and we can deduce a phase diagram by calculating the Chern numbers in the respective $S_z$ spin sectors. As a function of $M/B$ this leads to three phases.
For values of $M/B<8$ there is the $\Gamma$-phase which is the standard QSHI from the AII symmetry class with a $Z_2$ topological index $\nu=1$~\cite{kane2005}. This phase can also be understood as a stack of two time-reversed Chern insulators of Chern number~\cite{tknn,haldane} $\mathcal{C}_\uparrow=-\mathcal{C}_\downarrow= -1$, which leads to the pair of helical boundary modes related to each other by time-reversal symmetry. For values $8<M/B<12$ we find the X-Y valley phase, {indexed as $p4$ in Ref.\ \onlinecite{slager2013}}, which we refer to as PGI. In this parameter regime the time-reversed bands have Chern number $\mathcal{C}_\uparrow=-\mathcal{C}_\downarrow= -2$. This results in two pairs of helical edge modes, where each of the pairs consists of counterpropagating modes related by time-reversal symmetry. This phase is trivial in the sense of the periodic table of electronic topological states, i.e.,  the $\rm{Z}_2$-index is trivial ($\nu=0$), but features a non-trivial topological invariant $C_R$ stemming from the $C_4$ symmetry, as shown above. For higher values of $M/B$ the model enters a trivial phase, meaning that both the spin  Chern numbers as well as the $\rm{Z}_2$-index are trivial ($\mathcal{C}_{\uparrow,\downarrow}=0$, $\nu=0$, and $C_R=0$). All phases possess time-reversal symmetry and are connected via quantum phase transitions with quantum critical points that are metallic.

\begin{center}
\begin{figure}
\includegraphics[width=0.48\textwidth]{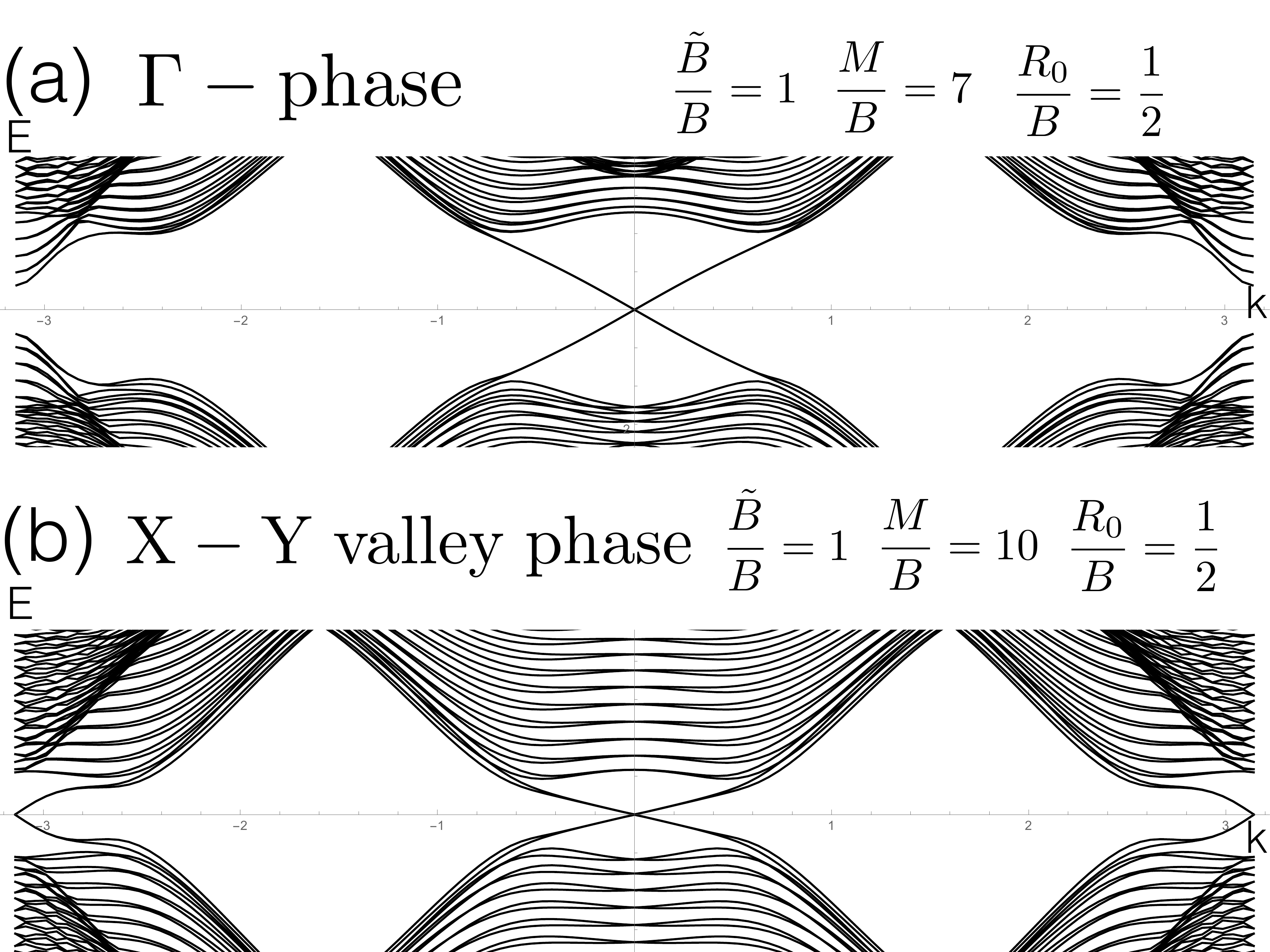}
\caption{Finite size spectrum on a cylinder geometry in the $\Gamma$-phase for parameters indicated in the inset. (a) In the $\Gamma$-phase one observes one set of helical modes at each boundary. (b) In the X-Y valley phase we find two pairs of helical modes at each boundary.}\label{fig:finitesize}
\end{figure}
\end{center}

Upon turning on the Rashba spin orbit coupling,  $R_0 \neq 0$, all symmetries of the system are preserved, except for spin-rotational. Therefore, we cannot define the spin Chern numbers any more (the $Z_2$-index is still well defined and calculable, though), but we can still define $C_R$. At a critical value of the Rashba spin-orbit coupling $R_{0c}/B$, the bulk band-gap closes, and all the phases become metallic as $R_0$ is further increased. To show that the clean system still features edge states without conserved spin, we studied the finite size spectrum on a cylinder. In the case of the $\Gamma$-phase for finite $R_0<R_{0c}$ we find one pair of helical edge states (Fig.~\ref{fig:finitesize} (a)) while in the X-Y valley phase we find two pairs (except for diagonal cut), see Fig.~\ref{fig:finitesize} (b).

Overall we conclude that finite Rashba coupling does not change the system properties in the clean system in either phase below a respective critical $R_{0c}$ where the insulator is converted into a metal.

\section{Transport properties}

One of the most striking properties of topological insulator states is not only the existence of boundary modes but also their stability with respect to disorder. The most prominent example is the quantum Hall state which has quantized Hall conductivity  $\sigma_H=\frac{e^2}{h}n$, with $n$ as an integer. The integer $n$ can be viewed in two equivalent ways: (i) it is the cumulative Chern number of the bands below the chemical potential or (ii) the number of chiral boundary modes. The conductance then accounts for the number of channels at the boundary. Naively, one would expect disorder to localize these modes, but their chiral nature provides an escape route: disorder cannot localize them since there is no way of converting a left-mover into a right-mover and vice versa by virtue of them being unidirectional.

\begin{figure}[h]
\includegraphics[width=0.46\textwidth]{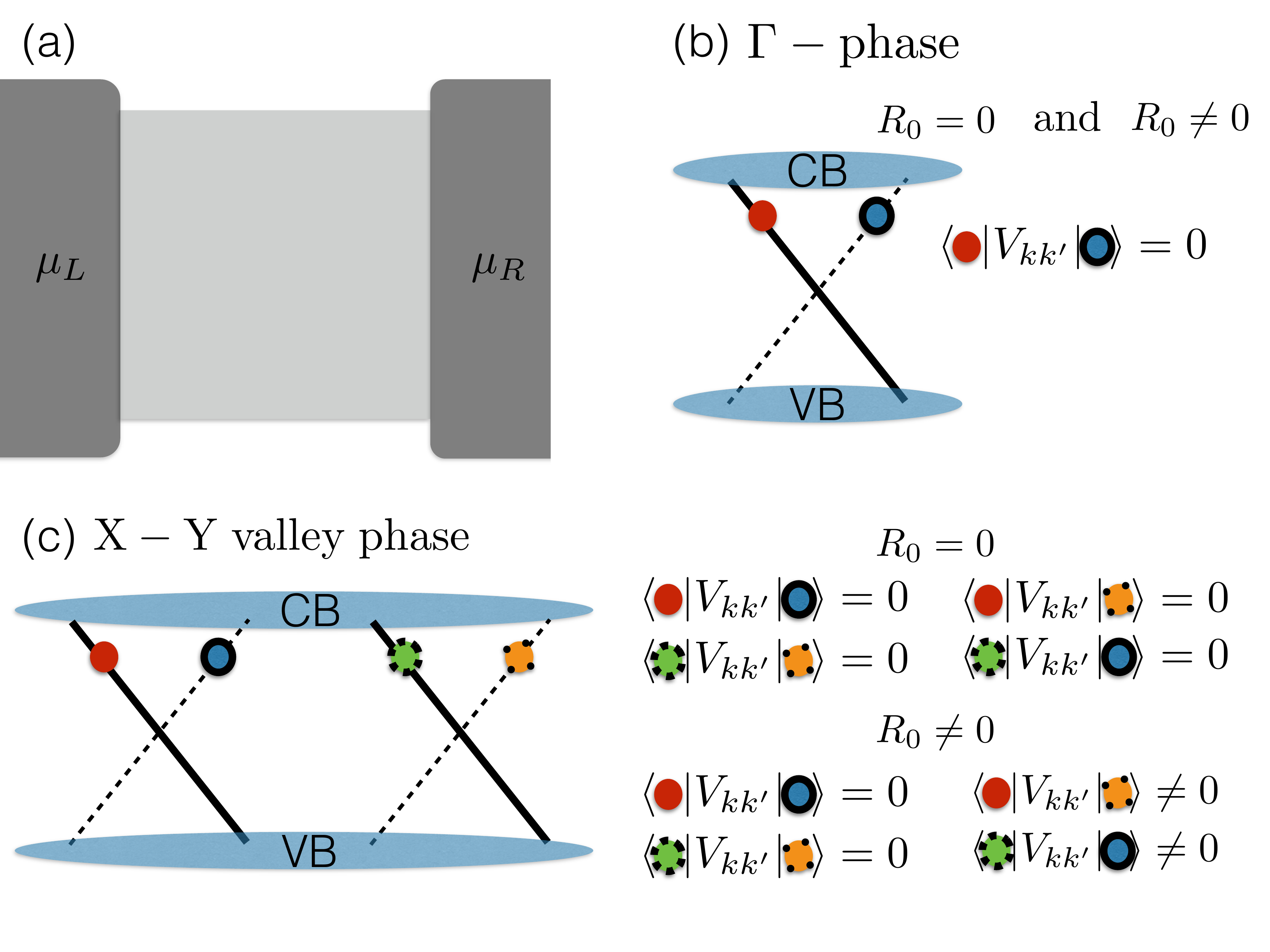}
\caption{ (a) Experimental setup to probe the transport properties of a sample sandwiched between two leads kept at different chemical potentials $\mu_L$ and $\mu_R$. (b) Counterpropagating modes at the boundary of a QSHI. Left- and right-movers are Kramers' partners and are protected against backscattering both for $R_0=0$ and $R_0 \neq 0$. $V_{kk'}=V_0$ denotes a featureless elastic scatterer which mixes all momenta. The zero overlap between the counterpropagating modes stems from additional quantum numbers of the band structure. (c) Two sets of counterpropagating modes at the boundary of a PGI. While the modes crossing at the time-reversal invariant momenta cannot scatter into each other due to time-reversal symmetry for both $R_0=0$ and $R_0 \neq 0$, modes from different time-reversal invariant momenta can scatter if $R_0 \neq 0$.}\label{fig:setup}
\end{figure}

In the case of a QSHI we do not have chiral channels but instead helical modes. This implies that there are right- and left-movers on either side of the sample which potentially allows for elastic backscattering due to scalar disorder. However, the helical modes are Kramers' pairs related by time-reversal symmetry, which implies that scalar disorder cannot convert left-movers into right-movers and vice versa due to the orthogonality of Kramers' pairs. The situation is schematically depicted in Fig.~\ref{fig:setup} (b). Consequently, edge transport is also ballistic resulting in a quantized conductance. This has also been observed in transport experiments on HgTe/CdTe quantum wells.~\cite{koning2006}

In the case of the PGI we find there are two pairs of boundary modes at each sample edge, see right-hand-side of Fig.~\ref{fig:finitesize} (b), but usually two pairs of modes are unstable since any right-mover in general is not mutually orthogonal to both left-movers and they can thus scatter into each other under generic circumstances. So the question we address here is in which sense and under which conditions the $X-Y$ valley phase, topologically trivial according to the periodic table, has stable transport properties once disorder is added.

The setup that we probe is shown in Fig.~\ref{fig:setup} (a) where we connect a central sample region to left and right leads and measure the response coefficient for transport from the left lead to the right lead in the linear response regime (we assume $\mu_L=\mu_R+\delta V$, $\delta V$ being small).

In order to study the stability of the edge states with respect to disorder we resort to the well established non-equilibrium Green-function method, which we implement numerically in the linear response regime~\cite{dattabook}.

In the clean system we can count the boundary modes assuming ballistic transport based on the finite size spectra (see Fig.~\ref{fig:finitesize}), and find
\begin{eqnarray}\label{eq:conductance}
G=\frac{e^2}{h} \times \left \{ \begin{array}{cc} 2 & {\rm{in}} \;\Gamma-{\rm{phase}} \\ 4 & {\rm{in}}\;{\rm{X-Y \; valley \; phase}} \end{array} \right. \;.
\end{eqnarray}
(Note that the result in the $X-Y$ phase does not hold if the edge is cut along the diagonal)

We have modelled disorder by scalar disorder, i.e., through local variations in the chemical potential. We have chosen disorder of the box type with a strength $w$, thus the local energy is drawn from $[-w,w]$. This preserves all the symmetries of the system required by the classification scheme, but it obviously breaks translational and rotational symmetries (locally, but not on average).
Subsequently, we have checked the localization tendencies by probing the system (i) with a fixed sample size as a function of increasing disorder strength, and (ii) at fixed disorder strength as a function of increasing sample size.

In order to make better comparisons between the QSHI and PGI , we have chosen parameters such that bulk gaps are approximately the same in both systems.  We found that a convenient parameter set is given by $\tilde{B}=B$,  $M/B=7$ (QSHI) or $M/B=10$ (PGI) without Rashba coupling, while with Rashba coupling  $R_0/B=1$ we choose $M/B=9$ for the PGI (to have comparable gaps).

\begin{figure}[h]
\includegraphics[width=0.48\textwidth]{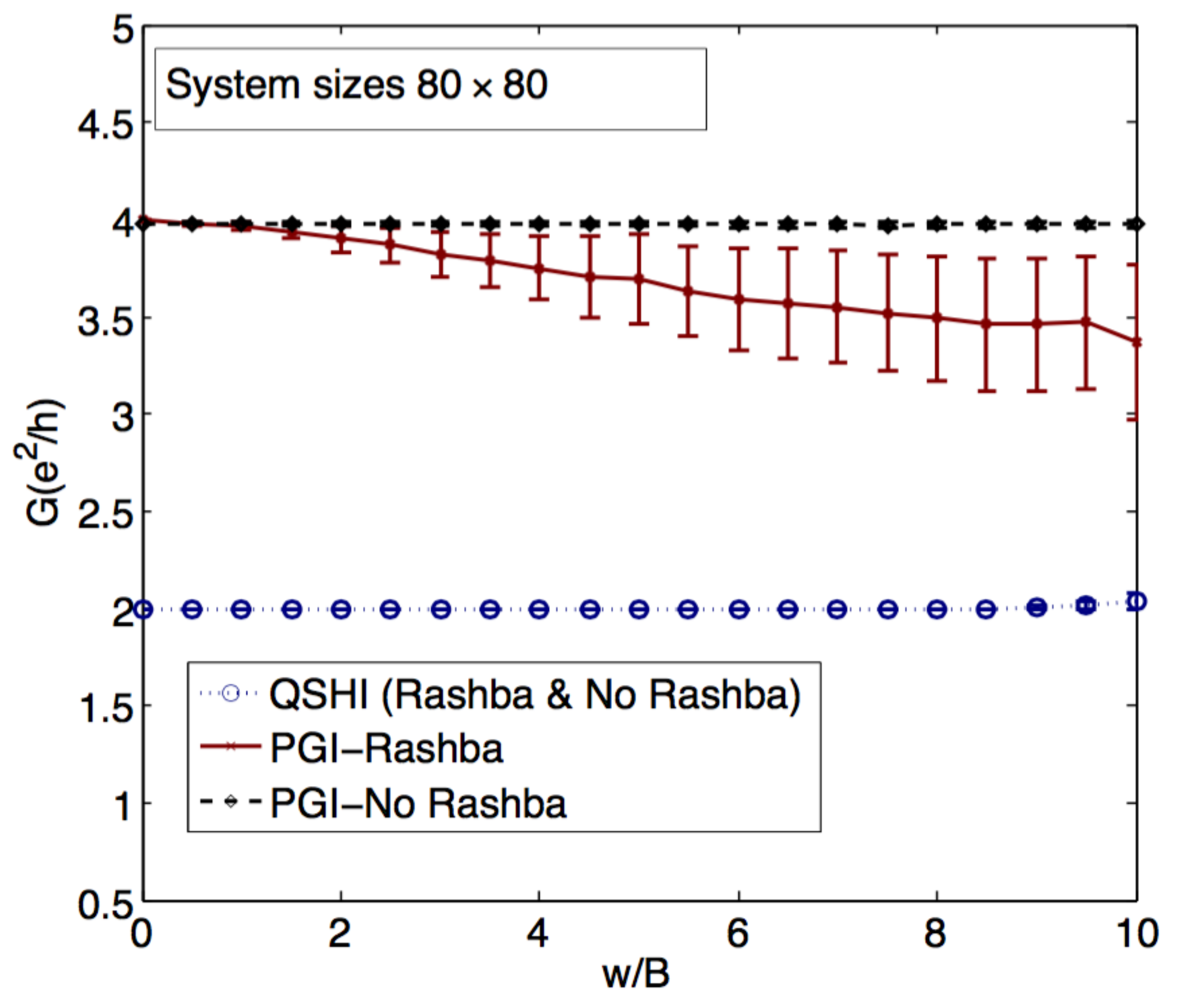}
\caption{Conductance of finite size system as a function of increasing disorder strength. In the case where the $S_z$-component of the spin is a conserved quantity both the QSHI and PGI are equally stable. When the Rashba spin-orbit coupling is switched on, the QSHI is still equally stable, while the protection for the PGI is lost. }\label{fig:finitesizevarydis}
\end{figure}

\begin{figure}
\includegraphics[width=0.48\textwidth]{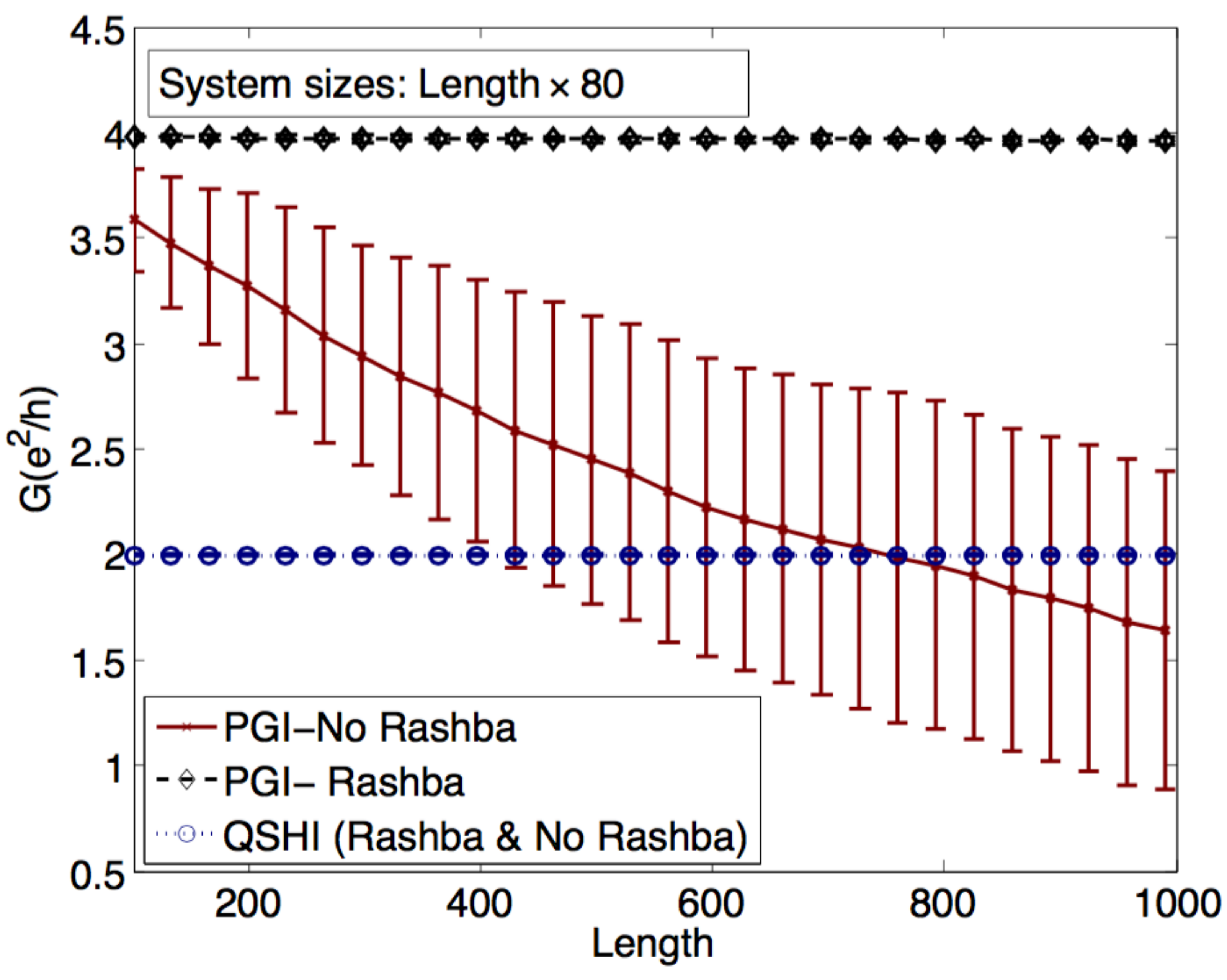}
\caption{Conductance of finite size systems at fixed disorder for increasing system sizes. For $R_0=0$ we find that both QSHI as well as PGI are equally stable, while for finite $R_0$ the conductance of the PGI decreases as a function of the system size.}\label{fig:fixeddisvarysize}
\end{figure}

In the absence of Rashba spin orbit coupling ($R_0=0$) with conserved spin, we find that both  $\Gamma$ and the X-Y valley phase are equally stable against disorder. This is not very surprising since both in the $\Gamma$-phase as well as in the X-Y valley phase we can think of the system as two time-reversed Chern insulators with Chern numbers $\mathcal{C}_{\uparrow,\downarrow}=\mp 1$ for the QSHI and $\mathcal{C}_{\uparrow,\downarrow}=\mp 2$ for the PGI and the left- and right-moving channels do not mix, see Fig.~\ref{fig:setup} (b) and (c). In Fig.~\ref{fig:finitesizevarydis}, we display a plot of the conductance of a sample of size $80\times 80$ with the chemical potential in the bulk gap as a function of disorder strength. Our results show that both the QSHI and the PGI phases are   stable against disorder, since the conductance is quantized. In Fig.~\ref{fig:fixeddisvarysize} we   plot conductance at fixed disorder strength as a function of the system size (we always study systems of transverse size 80 sites), and find again that both QSHI and PGI are stable.

 In the presence of Rashba spin-orbit coupling ($R_0\neq0$), we find that the features of the QSHI phase are unchanged, as expected. However, the protection of the conductance is lost in the case of the PGI. This can explicitly seen in Fig.~\ref{fig:finitesizevarydis}, where the conductance as a function of disorder strength at fixed sample size decreases, as soon as Rashba spin orbit coupling is switched on.  Furthermore, in Fig.~\ref{fig:fixeddisvarysize} we observe that the conductance as a function of the system size decreases and eventually would vanish if we made the sample long enough. This signals the localization of the modes in agreement with the absence of topological protection. As discussed before, this can be traced back to the loss of orthogonality of left- and right-movers at the sample edges. We have explicitly checked the loss of orthogonality under scattering from a structurless impurity ($V_{kk'}=V_0$) of the left- and right-movers in the finite size spectrum (as also discussed in Fig.~\ref{fig:setup} (c)). We can therefore conclude that the absence of protection against backscattering leads to the localization of the boundary modes.

\section{Conclusions}
In this paper we have studied the stability of the edge states of a topological system outside the tenfold classification of the topological insulator states with respect to disorder. We considered a special instance of a PGI where discrete rotational $C_4$ symmetry guarantees the existence of edge states in a clean system. If the $S_z$-component of the spin is a conserved quantity in this system we find that the boundary modes are protected against localization due to disorder. This protection against disorder is lost if Rashba spin-orbit coupling is present. Consequently, the system indeed behaves like a trivial insulator in the sense of the topological classification of electronic states. For the future it is an interesting prospect to study the localization properties of other systems with boundary modes which are outside the tenfold classification of topological states, such as the recently observed three-dimensional TCI phase featuring band-inversions at symmetry-related $L-$points in the Brillouin zone.\cite{ando2012,xu2012,story2012}

\section{Acknowledgement}
We thank Michael Wimmer and Jason Frank for helpful discussions. We acknowledge funding from the DFG FR 2627/3-1 (C.K. and L.F.). This work is part of the D-ITP consortium, a program of the Netherlands Organisation for Scientific Research (NWO) that is funded by the Dutch Ministry of Education, Culture and Science (OCW). V. J. acknowledges financial support from NWO.

\appendix

\section{Invariance of the Hamiltonian under $C_4$ rotation}

In this Appendix we explicitly show that the Hamiltonian in Eq.\ (\ref{eq:H}) in the main text is invariant under $C_4$ rotations.
$C_4$ rotation is represented by
\begin{equation}
{\cal R}={\cal R}_{\rm orb}\otimes{\cal R}_{\rm spin},
\end{equation}
with
\begin{eqnarray}
&{\cal R}_{\rm orb}=\frac{1}{2}[(1+i)\tau_0+(1-i)\tau_z],\\
&{\cal R}_{\rm spin}={\exp}(i\frac{\pi}{4}\sigma_z)=\frac{1}{\sqrt{2}}(1+i\sigma_z).
\end{eqnarray}
This symmetry operation acts on the components of the momentum as
\begin{equation}\label{C4momemtum}
k_{x,y}\rightarrow \pm k_{y,x}.
\end{equation}

First, the Hamiltonian (\ref{eq:H}) in the main text without the Rashba term ($R_0=0$) can be conveniently rewritten as
\begin{equation}
H_0= d_x({\bf k})\tau_x\otimes\sigma_z+(d_y({\bf k})\tau_y+d_z({\bf k})\tau_z)\otimes\sigma_0,
\end{equation}
and is clearly invariant under $C_4$ spin rotation. The invariance under $C_4$ rotation in the orbital space is
a consequence of the fact that
\begin{equation}\label{C4orb}
{\cal R}_{\rm orb}^\dagger \tau_{x,y}{\cal R}_{\rm orb}=\mp\tau_{y,x},
\end{equation}
and the transformation of the momentum under $C_4$ given by Eq.\ (\ref{C4momemtum}).
Namely, the form of the $C_4$ transformation yields
$d_{x,y}({\bf k})\rightarrow\mp d_{y,x}({\bf k}) $, while $d_z({\bf k})$ is invariant, which together with Eq.\ (\ref{C4orb}),
implies the invariance of $H_0$ under the $C_4$ rotation.

To show that $C_4$ is a symmetry of the Rashba Hamiltonian, we conveniently rewrite it as
\begin{equation}\label{HR}
H_R=\frac{1}{2}R_0(\tau_0+\tau_z)\otimes(\sigma_y\sin k_x-\sigma_x\sin k_y).
\end{equation}
The orbital part of this Hamiltonian is  invariant under $C_4$ rotation, since the matrices $\tau_0$ and $\tau_z$ commute with each other.
Under $C_4$ rotation, Pauli matrices $\sigma_{x,y}$ transform as
\begin{equation}\label{C4spin}
{\cal R}_{\rm spin}^\dagger\sigma_{x,y} {\cal R}_{\rm spin}=\pm \sigma_{y,x}.
\end{equation}
Transformation of the momentum under $C_4$, Eq.\ (\ref{C4momemtum}), implies that $\sin k_{x,y}\rightarrow \pm \sin k_{y,x}$ under the same operation. This, together with Eq.\ \ref{C4spin}, implies
the invariance of the spin part of the Hamiltonian (\ref{HR}), which shows
the invariance of the Hamiltonian (\ref{HR}) under $C_4$ transformation.


\begin{thebibliography}{10}

 \bibitem{kane2005}
C. L. Kane, E. J. Mele,  Phys. Rev. Lett. {\bf 95},
  146802 (2005).

\bibitem{kane2005a}
C.\ L.\ Kane, E. J. Mele,  Phys. Rev. Lett. {\bf 95}, 226801 (2005).

 \bibitem{bernevigzhang2006}
B. A. Bernevig, T.L. Hughes, S.C. Zhang, Science {\bf 314},
 1757 (2006).

 \bibitem{koning2006}
 M.~K\"onig, {\it et~al.\/}, Science {\bf 318}, 766 (2007).




  \bibitem{moore2007}
  J.E. Moore, L. Balents, Phys. Rev. B {\bf 75},
121306 (2007).

\bibitem{fukane2006}
L. Fu, C. L. Kane, Phys. Rev. B {\bf74},
 195312 (2006).

 \bibitem{fukane2007a}
L. Fu, C. L. Kane, Phys. Rev. Lett. {\bf 98},
  106803 (2007).
  \bibitem{fukane2007b}
L. Fu, C. L. Kane, Phys. Rev. B {\bf 76},
  045302 (2007).

\bibitem{hsieh2008}
 D.~Hsieh, {\it et~al.\/},   Nature {\bf 452}, 970 (2008).

\bibitem{hsieh2009}
 D.~Hsieh, {\it et~al.\/},   Science {\bf 323}, 919 (2009).

 \bibitem{xia2009}
Y.~ Xia, {\it et~al.\/},  Nature Phys. {\bf 5}, 398 (2009).


\bibitem{zhang2009}
H.~ Zhang, {\it et~al.\/},  Nature Phys. {\bf 5}, 438 (2009).

\bibitem{chen2009}
Y. L. ~Chen, {\it et~al.\/},   Science {\bf 325}, 178 (2009).


\bibitem{hasan-kane-review}
\textrm{M.\ Z.\ Hasan}, \textrm{C.\ L.\ Kane},   Rev. Mod. Phys. {\bf 82}, 3045 (2010).

\bibitem{qi-zhang-review}
\textrm{X.\ L.\ Qi}, \textrm{S.\ C.\ Zhang}, Rev. Mod. Phys. {\bf 83}, 1057 (2011).


  \bibitem{schnyder2008}
\textrm{A. P. Schnyder}, \textrm{ S. Ryu},\textrm{  A. Furusaki}, \textrm{A.W.W. Ludwig},
Phys. Rev. B {\bf 78}, 195125 (2008).



 \bibitem{kitaev2009}
 \textrm{A. Kitaev},    AIP Conf. Proc. {\bf 22},
  1132 (2009).

\bibitem{Qi2008}{
X.-L.~Qi, T.~Hughes, S.-C.~Zhang, Phys.\ Rev.\ B \textbf{78}, 195424 (2008).
}

\bibitem{schnyderNJP2010}
\textrm{S. Ryu, A. P. Schnyder, A. Furusaki, A. Ludwig},
New J. Phys., {\bf 12}, 065010 (2010).



\bibitem{fu2011}
L. Fu, Phys. Rev. Lett. {\bf 106},
106802 (2011).

\bibitem{slager2013}
R.-J. Slager, A. Mesaros, V. Juri\v ci\' c, and J. Zaanen, Nature Phys. {\bf 9}, 98 (2013).

\bibitem{chiu2013}
C.-K.\ Chiu, H.\ Yao, and S.\ Ryu, Phys.\ Rev.\ B {\bf 88}, 075142 (2013).

\bibitem{morimoto2013}
T.\ Morimoto and A.\ Furusaki, Phys.\ Rev.\ B {\bf 88}, 125129 (2013).

\bibitem{shiozaki2014}
K.\ Shiozaki and M.\ Sato, Phys.\ Rev.\ B {\bf 90}, 165114 (2014).

\bibitem{fuNatComm2012}
T.\ H.\ Hsieh, H. Lin, J. Liu, W. Duan, A. Bansil, L. Fu, Nat. Comm. {\bf 3}, 982 (2012)

\bibitem{ando2012}
Y.~ Tanaka,	Z.\ Ren, T.\ Sato, K.\ Nakayama, S. Souma, T.\ Takahashi, K.\ Segawa, and Y.\ Ando, Nature Phys. {\bf 8}, 800 (2012).

\bibitem{xu2012}
S.-Y.\ Xu, C.\ Liu, N.\ Alidoust, M.\ Neupane, D.\ Qian, I.\ Belopolski, J.\ D.\ Denlinger, Y.\ J.\ Wang, H.\ Lin, L.\ A.\ Wray, G.\ Landolt,
B.\ Slomski, J.\ H.\ Dil, A.\ Marcinkova, E.\ Morosan, Q.\ Gibson, R.\ Sankar, F.\ C.\ Chou, R.\ J.\ Cava, A.\ Bansil, and M.\ Z.\ Hasan,  { Nature Commun.} {\bf 3}, 1192  (2012).

\bibitem{story2012}
P.\ Dziawa, B.\ J.\ Kowalski, K.\ Dybko, R.\ Buczko, A.\ Szczerbakow, M.\ Szot, E.\ Lusakowska, T.\ Balasubramanian, B.\ M.\ Wojek, M.\ H.\ Berntsen,
O.\ Tjernberg, and T.\ Story, { Nature Mater.} {\bf 11}, 1023  (2012).

\bibitem{tanaka2013}
Y.\ Tanaka, T.\ Sato, K.\ Nakayama, S.\ Souma, T.\ Takahashi, Z.\ Ren, M.\ Novak, K.\ Segawa, and Y.\ Ando, Phys. Rev. B {\bf 87}, 155105 (2013).

\bibitem{okada2014}
Y.\ Okada, M.\ Serbyn, H.\ Lin, D.\ Walkup, W.\ Zhou, C.\ Dhital, M.\ Neupane, S.\ Xu, Y.\ J.\ Wang, R.\ Sankar, F.\ Chou, A.\ Bansil, M.\ Z.\ Hasan, S.\ D.\ Wilson, L.\ Fu and V.\ Madhavan, Science {\bf 341}, 6153 (2014).

\bibitem{zeljkovic2014}
 I.\ Zeljkovic, Y.\ Okada, C.\ Y.\ Huang, R.\ Sankar, D.\ Walkup, W.\ Zhou, M.\ Serbyn, F.\ Chou, W.\ F.\ Tsai, H.\ Lin, A.\ Bansil, L.\ Fu, M.\ Z.\ Hasan, and V.\ Madhavan, Nature Phys. {\bf 10}, 572 (2014).

 \bibitem{jiang2014}
 H.\ Jiang, H.\ Liu, J.\ Feng, Q.\ Sun, and X.\ C.\ Xie, Phys. Rev. Lett. {\bf 112}, 176601 (2014).

\bibitem{ezawa2014}
M. Ezawa, New J. Phys. {\bf 16}, 065015 (2014).

 \bibitem{rothe2010}
D.\ G.\ Rothe, R.\ W.\ Reinthaler, C.-X.\ Liu, L.\ W.\ Molenkamp, S.-C.\ Zhang, and E.\ M.\ Hankiewicz, New J. Phys. {\bf 12}, 065012 (2010).

\bibitem{footnote}
An analogous derivation can be carried out based on the results in Ref.\ \onlinecite{fang2012}.

\bibitem{fang2012}
C.\ Fang, M.\ J.\ Gilbert, and B.\ A.\ Bernevig, Phys.\ Rev.\ B {\bf 86}, 115112 (2012).

\bibitem{fu-surface}
J.\ Liu, W.\ Duan, and L.\ Fu, Phys. Rev. B {\bf 88}, 241303(R) (2013).

\bibitem{tknn}
D. J. Thouless, M. Kohmoto, M. P. Nightingale, M. den Nijs,
 Phys. Rev. Lett. {\bf 49}, 405 (1982).

\bibitem{haldane}
F.~D.~M.~Haldane, Phys.\ Rev.\ Lett.\ {\bf{61}}, 2015 (1988).

\bibitem{dattabook}
S. Datta, {\it Electronic Transport in Mesoscopic Systems}, Cambridge University Press (1997).


\end{thebibliography}
\end{document}